\documentclass[prl,twocolumn,letterpaper,superscriptaddress,longbibliography]{revtex4-1}

\usepackage{graphicx}
\usepackage{dcolumn}
\usepackage{bm}
\usepackage{color}
\usepackage{amsmath}
\usepackage{stmaryrd}

\begin{document}

\title{Tunable Magnon-Photon Coupling by Magnon Band Gap in a Layered Hybrid Perovskite Antiferromagnet}

\author{Yi Li}
\email{yili@anl.gov}
\affiliation{Materials Science Division, Argonne National Laboratory, Argonne, IL 60439, USA}

\author{Timothy Draher}
\affiliation{Materials Science Division, Argonne National Laboratory, Argonne, IL 60439, USA}
\affiliation{Northern Illinois University, Department of Physics, Dekalb Illinois, 60115, USA}

\author{Andrew H. Comstock}
\affiliation{Department of Physics and Organic and Carbon Electronics Laboratory (ORACEL), North Carolina State University, Raleigh, NC 27695 USA}

\author{Yuzan Xiong}
\affiliation{Department of Physics and Astronomy, University of North Carolina, Chapel Hill, NC 27599, USA}
\affiliation{Materials Science Division, Argonne National Laboratory, Argonne, IL 60439, USA}

\author{Md Azimul Haque}
\affiliation{Chemistry and Nanoscience Center, National Renewable Energy Laboratory, Golden, CO 80401, USA}

\author{Elham Easy}
\affiliation{Department of Mechanical Engineering, Stevens Institute of Technology, Hoboken, NJ 07030, USA}

\author{Jiang-Chao Qian}
\affiliation{Department of Materials Science and Engineering, University of Illinois Urbana-Champaign, Urbana, IL 61820, USA}

\author{Tomas Polakovic}
\affiliation{Physics Division, Argonne National Laboratory, Lemont, IL 60439, USA}

\author{John E. Pearson}
\affiliation{Materials Science Division, Argonne National Laboratory, Argonne, IL 60439, USA}

\author{Ralu Divan}
\affiliation{Center for Nanoscale Materials, Argonne National Laboratory, Argonne, IL 60439, USA}

\author{Jian-Min Zuo}
\affiliation{Department of Materials Science and Engineering, University of Illinois Urbana-Champaign, Urbana, IL 61820, USA}

\author{Xian Zhang}
\affiliation{Department of Mechanical Engineering, Stevens Institute of Technology, Hoboken, NJ 07030, USA}

\author{Ulrich Welp}
\affiliation{Materials Science Division, Argonne National Laboratory, Argonne, IL 60439, USA}

\author{Wai-Kwong Kwok}
\affiliation{Materials Science Division, Argonne National Laboratory, Argonne, IL 60439, USA}

\author{Axel Hoffmann}
\affiliation{Department of Materials Science and Engineering, University of Illinois Urbana-Champaign, Urbana, IL 61820, USA}

\author{Joseph M. Luther}
\affiliation{Chemistry and Nanoscience Center, National Renewable Energy Laboratory, Golden, CO 80401, USA}

\author{Matthew C. Beard}
\affiliation{Chemistry and Nanoscience Center, National Renewable Energy Laboratory, Golden, CO 80401, USA}

\author{Dali Sun}
\email{dsun4@ncsu.edu}
\affiliation{Department of Physics and Organic and Carbon Electronics Laboratory (ORACEL), North Carolina State University, Raleigh, NC 27695 USA}

\author{Wei Zhang}
\email{zhwei@unc.edu}
\affiliation{Department of Physics and Astronomy, University of North Carolina, Chapel Hill, NC 27599, USA}

\author{Valentine Novosad}
\email{novosad@anl.gov}
\affiliation{Materials Science Division, Argonne National Laboratory, Argonne, IL 60439, USA}

\date{\today}

\begin{abstract}
Tunability of coherent coupling between fundamental excitations is an important prerequisite for expanding their functionality in hybrid quantum systems. In hybrid magnonics, the dipolar interaction between magnon and photon usually persists and cannot be switched off. Here, we demonstrate this capability by coupling a superconducting resonator to a layered hybrid perovskite antiferromagnet, which exhibits a magnon band gap due to its intrinsic Dzyaloshinskii-Moriya interaction. The pronounced temperature sensitivity of the magnon band gap location allows us to set the photon mode within the gap and to disable magnon-photon hybridization. \textcolor{black}{When the resonator mode falls into the magnon band gap, the resonator damping rate increases due to the nonzero coupling to the detuned magnon mode. This phenomena can be used to quantify the magnon band gap using an analytical model.} Our work brings new opportunities in controlling coherent information processing with quantum properties in complex magnetic materials.
\end{abstract}

\maketitle

Hybrid quantum systems \cite{KurizkiPNAS2015,ClerkNPhys2020,LachanceQuirionAPEx2019} offer an important pathway for harnessing different natural advantages of complementary quantum systems, leveraging the distinct properties of their constituent excitations. The fundamental excitations of magnetically ordered materials, i.e. magnons, provide efficient coupling with other excitations \cite{LiJAP2020,LiAPLMater21}, such as microwave photons \cite{HueblPRL2013,GoryachevPRApplied14,TabuchiPRL2014,ZhangPRL2014,BaiPRL2015,LiPRL2019_magnon,HouPRL2019}, acoustic phonons \cite{KikkawaPRL16,BerkNComm19,AnPRB20,XuPRApplied21}, and magnons themselves \cite{KlinglerPRL18,ChenPRL18,MacNeillPRL19,LiPRL20,ShiotaPRL20,SudPRB20,SklenarPRApplied21,LimeiPRB21}, therefore holding promise for future integration with diverse quantum modules \cite{HarderSSP18,AwschalomIEEETQE21,FukamiPRXQ21,YuanPhysRep22}. In addition, coherent magnon interactions exhibit great controllability in different aspects, such as polarization \cite{LiPRL2019_magnon}, mode profile \cite{ShiotaPRL20}, phase \cite{HarderPRL2018,BhoiPRB2019,WangPRL19,BoventerPRR2020} and layer structure \cite{LiPRL20}, allowing for implementation of coherent magnon operations \cite{XuPRL21}. Recent demonstrations of coherent magnon-magnon coupling with controllable coupling strength by frequency detuning \cite{LambertPRA2016,LiPRL22} have further expanded the capability of distributed hybrid magnonic networks \cite{LiJAP2020}.

\begin{figure}[htb]
 \centering
 \includegraphics[width=3.2 in, angle = 0]{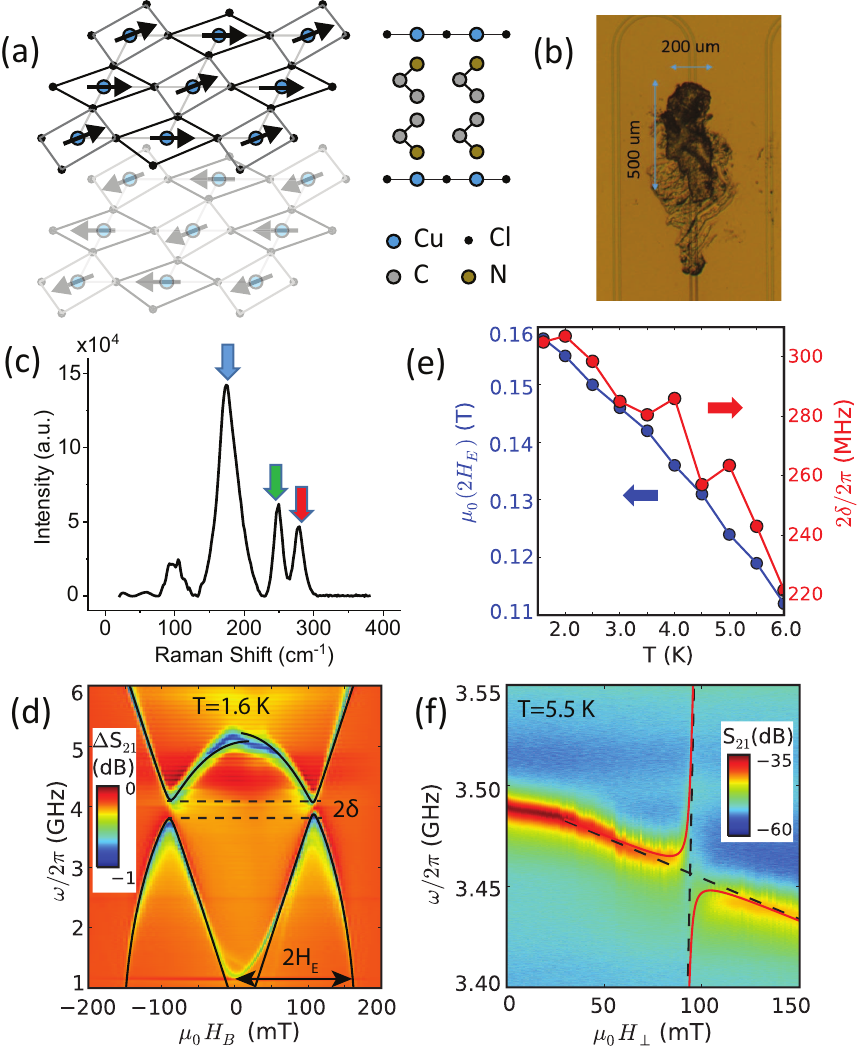}
 \caption{(a) Lattice structure of layered perovskite antiferromagnet Cu-EA, with Cu filling the octahedral sites of Cl and the antiferromagnetic layers separated by the CH$_3$CH$_2$NH$_3$ molecules. (b) Optical microscope image of a small Cu-EA flake mounted onto a CPW superconducting resonator. (c) Raman spectroscopy of Cu-EA showing the high-frequency octahedral modes and the low-frequency organic structure modes. (d) Broad-band ferromagnetic resonance spectra of a large Cu-EA crystal measured at 1.6 K, which is used to extract the magnon band gap $2\delta$, the interlayer exchange field $2H_E$, and magnon damping rate $\kappa_m$. \textcolor{black}{The colorbar shows the signals $\Delta S_{21}$ after background subtraction.} (e) Extracted $2H_E$ \text{magenta}{and $2\delta$} as a function of $T$. (f) Mode anticrossing between magnons and photons at 5.5 K, with $H_B\perp h_\text{rf}^y$. \textcolor{black}{The colorbar shows the signals $S_{21}$ in absolute values.} The red curves are the fits with $g/2\pi=45$ MHz. The black dashed lines denote the magnon and photon modes without interaction.}
 \label{fig1}
\end{figure}

\begin{figure*}[htb]
 \centering
 \includegraphics[width=6.0 in, angle = 0]{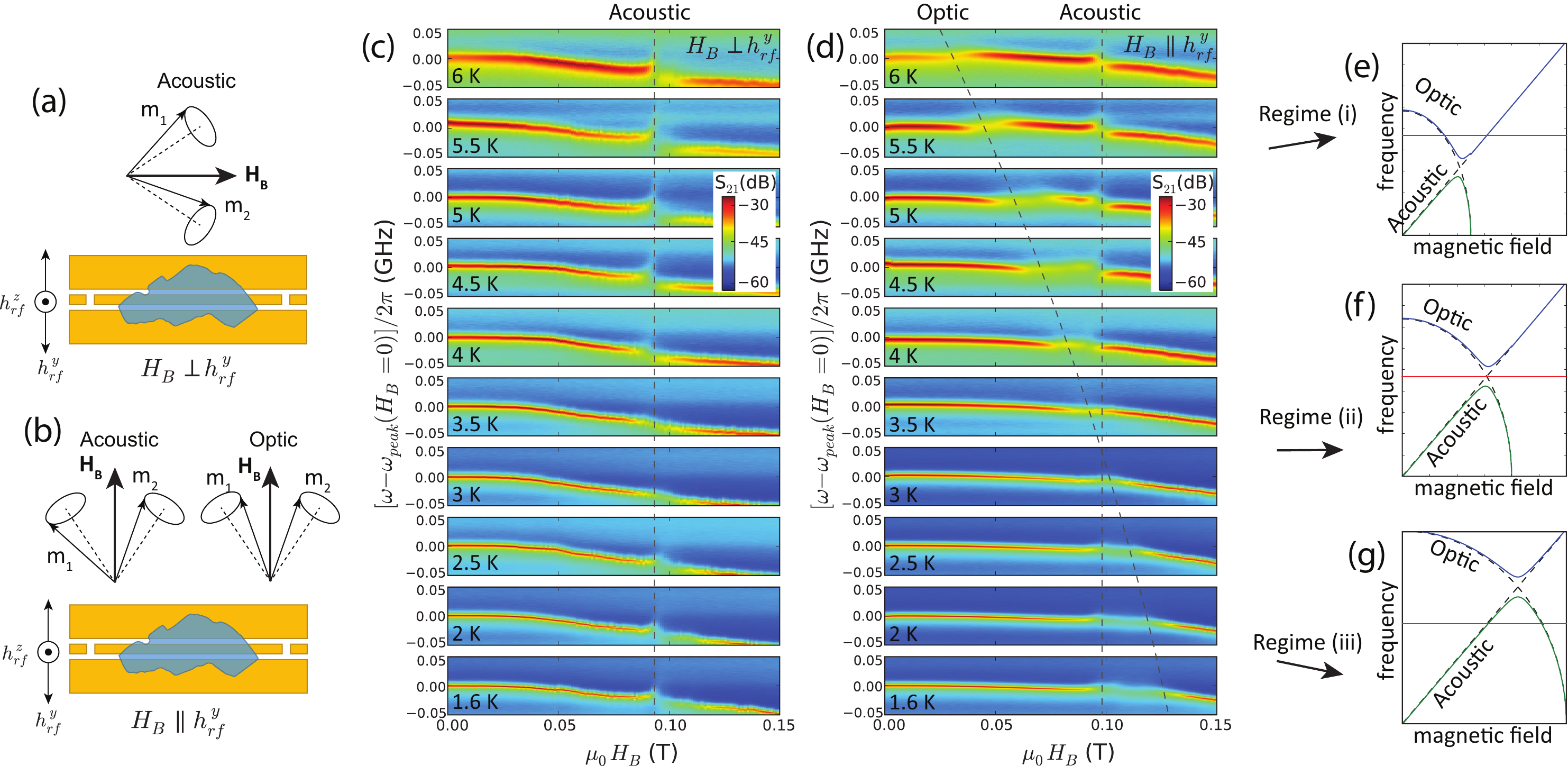}
 \caption{(a-b) Illustration of two different in-plane field alignments and their selective mode excitations. In (a), $\mu_{0}H_B\perp h_\textrm{rf}^y$ and only the acoustic mode is excited. In (b), $\mu_{0}H_B\parallel h_\textrm{rf}^y$ and both the acoustic and optical modes are excited. (c-d) Temperature dependence of the magnon-photon coupling evolutions from 1.5 to 6 K with two different magnetic field alignments. All the dispersion centered at the resonator photon mode ($\omega_p/2\pi\approx 3.5$ GHz). Dashed curves are guide to eye for the acoustic mode crossing the resonator mode at $\mu_0H_a=95$ mT and the optical mode crossing the resonator mode at different fields. (e-g). Illustration of the three regimes where the resonator mode is (e) above, (f) within, and (g) below the magnon band gap.}
 \label{fig2}
\end{figure*}

Despite its controllability, which is largely based on extrinsic control of magnetic systems, the intrinsic magnetic properties are rarely explored for manipulating coherent magnon interactions. To date, landmark demonstrations of hybrid magnonics have been centered on the ferrimagnetic insulator yttrium iron garnet (YIG) \cite{HueblPRL2013,TabuchiPRL2014,ZhangPRL2014,BaiPRL2015,TabuchiScience2015} or metallic magnets such as NiFe \cite{LiPRL2019_magnon,HouPRL2019}. Their relatively simple and rigid chemical and magnetic structures limit the potential for developing highly tunable hybrid systems. On the other hand, the recent two-dimensional (2D) organic layered magnets \cite{WangACSNano22} offer distinct advantages in their structure-enabled topological chirality and symmetry breaking \cite{JinarXiv2021}. One nice class of materials is 2D magnetic hybrid organic-inorganic perovskites (HOIPs) possessing both superior structural versatility and long-range magnetic order \cite{SaparovChemRev16,NugrohoPRB16,KimSmall20,WillettEJIC12}. They usually exhibit an interlayer antiferromagnetic (AFM) coupling \cite{Jongh72}, inducing the acoustic and optical magnon modes \cite{KefferPR52,MacNeillPRL19} in the gigahertz (GHz) frequency range. In addition, the structural symmetry breaking leads to Dzyaloshinskii--Moriya interaction (DMI) \cite{MoriyaPR60}, causing a finite spin canting \cite{BloembergenAIP73,BogdanovPRB07} and creating an intrinsic magnon band gap where the acoustic and optical modes intersect \cite{ComstockNComm23}. This is fundamentally different from the magnon band gap induced by an external field \cite{DiederichNNano23}\cite{MacNeillPRL19,ShiotaPRL20,SudPRB20,SklenarPRApplied21,LimeiPRB21,MakiharaNComm21} in that the DMI has provided an intrinsic effective field for magnon-magnon coupling without the need of external field. Furthermore, the large sensitivity of the magnon band gap to small temperature change can lead to new opportunities of modulating coherent magnonic coupling.

\cite{ZollitscharXiv2022}
In this Letter, we report a hybrid magnonic system consisting of a 2D HOIPs, (CH$_3$CH$_2$NH$_3$)$_2$CuCl$_4$ (Cu-EA) \cite{ComstockNComm23}, coupled to a superconducting resonator. The high sensitivity of the superconducting resonator enables \textcolor{black}{coherent} magnon-photon coupling and avoided crossing with a small Cu-EA flake. By changing the temperature of the sample, the location of the DMI-induced magnon band gap can be adjusted so that the resonator photon mode completely falls into the gap and cancels the mode hybridization. At the non-hybridized state, the magnetic interaction with the resonator causes the resonator linewidth to broaden. Using our developed analytical model, the narrow-band linewidth broadening measurements can be used to extract the magnon band gap, which quantitatively agrees with the broad-band FMR measurements. Our results highlight the opportunity of manipulating coherent mode hybridization with new quantum materials and probing their complex magnonic dispersion with narrow-band microwave characterizations.

The chemical structure of Cu-EA features corner-sharing halogen (Cl) octahedra with the Cu atom situated at the center, as shown in Fig. \ref{fig1}(a). The canted inorganic CuCl$_{4}^{2-}$ octahedral structures allow for intralayer long-range magnetic order with superexchange Cu-Cl-Cu interactions, while the interlayer organic cations modulate the interlayer antiferromagnetic (AFM) coupling \cite{Jongh72}. Raman spectroscopy of the Cu-EA \cite{EasyACS2021} confirms the vibration modes of the octahedral structure (at 175, 250, and 280 cm$^{-1}$) and the organic cation (at 100 cm$^{-1}$) \cite{CarettaPRB14}, as shown in Fig. \ref{fig1}(c). We have also conducted Inductively Coupled Plasma (IPC) spectroscopy on the sample, showing accurate stochiometry of the elemental weight as compared with the chemical structure \cite{sm}.

\textcolor{black}{Fig. \ref{fig1}(d) shows the broad-band ferromagnetic resonance of a large Cu-EA crystal at 1.6 K at parallel pumping condition, i.e. $\mu_0H_B\parallel h_\textrm{rf}^y$ as illustrated in Fig. \ref{fig2}(b). Both the acoustic and optical modes are measured, which can be formulated as \cite{MacNeillPRL19}:}
\begin{equation}\label{eq0}
\omega_a = \mu_0\gamma \sqrt{2H_E(2H_E+M_{eff})} {H\over2H_E}
\end{equation}
\begin{equation}\label{eq1}
\omega_o = \mu_0\gamma \sqrt{2H_EM_{eff}\left(1-{H^2 \over 4H_E^2}\right)}
\end{equation}
\textcolor{black}{where $H_E$ is the interlayer exchange coupling field, $M_{eff}$ is the effective magnetization which contributes to the perpendicular demagnetization field, and $\gamma/2\pi=(g_{e}/2)\times 28$ GHz/T is the gyromagnetic ratio, with $g_{e}$ as the $g$-factor of the magnetization. Clear avoided crossing gaps between the two modes show the existence of magnon band gap around 4 GHz. The coupled magnon spectra can be fitted to the hybrid mode expression $\omega^{mm}_\pm = (\omega_a+\omega_o)/2\pm\sqrt{(\omega_a-\omega_o)^2/4+\delta^2}$, where $\delta$ is the magnon-magnon coupling strength. The fitting curves are plotted in Fig. \ref{fig1}(d). The extracted parameters are $\mu_0H_E=0.16$ T, $\mu_0M_{eff}=80$ mT, $g_{e}=2.3$, and $\delta/2\pi=150$ MHz. Note that the actual saturation magnetization of Cu-EA can be larger than $M_{eff}$ because the shape of the sample crystal is not a perfect two-dimension system and the perpendicular demagnetization factor can be smaller than one. The strong magnon-magnon coupling observed in Cu-EA at parallel pumping condition, which is absent in other layered \cite{MacNeillPRL19} or synthetic \cite{ShiotaPRL20,SudPRB20} antiferromagnets at the same pumping condition, is caused by the spontaneous canting of the octahedral CuCl$_4$$^{2-}$ spin sites from their chiral DMI and the resultant overlap between the acoustic and optical modes \cite{ComstockNComm23}.} \textcolor{black}{Fig. \ref{fig1}(e) shows the temperature dependence of extracted $2H_E$ and the magnon band gap $2\delta$ from Fig. \ref{fig1}(d). With the same $y$-axis proportion ratio in Fig. \ref{fig1}(e), a good overlap of $H_E$ and $\delta$ shows that they are proportional to each other at different temperatures. This suggests that the DMI-induced spin canting shares a similar mechanism with the interlayer exchange coupling in Cu-EA.} \textcolor{black}{The magnon damping rate, $\kappa_m$, of the acoustic and optical modes are also extracted and are found to be weakly frequency and temperature dependent. In the range of 2-5 GHz, $\kappa_m/2\pi\sim 50$ MHz for the acoustic mode and $\sim 80$ MHz for the optical mode; see the Supplemental Materials for details \cite{sm}.}

To feature the sensitivity of the superconducting resonator to small magnetic crystals, we precisely transfer a thin Cu-EA flake with lateral dimensions of 500 $\mu$m $\times$ 200 $\mu$m and a thickness of 40 $\mu$m onto the center of a half-wavelength NbN coplanar waveguide (CPW) superconducting resonator with a signal line width of 20 $\mu$m, as shown in Fig. \ref{fig1}(b). The dimension matching between the signal line and the flake thickness allows for optimal coupling of the magnon excitations to the resonator. The loaded superconducting resonator exhibits a sharp peak at $\omega_p/2\pi=3.5$ GHz and a zero-field half-width half-maximum linewidth of $\kappa_p/2\pi=0.4$ MHz at 1.6 K, which corresponds to a quality factor of $\omega_p/2\kappa_p=4400$.

The maximum mode splitting happens at 5.5 K between the acoustic magnon mode of Cu-EA and the resonator photon mode, shown in Fig. \ref{fig1}(f). The peak positions of the avoided crossing can be fitted to the hybrid modes \cite{HueblPRL2013}:
\begin{equation}\label{eq00}
\omega^{mp}_\pm = (\omega_m+\omega_p)/2\pm\sqrt{(\omega_m-\omega_p)^2/4+g^2}
\end{equation}
where $\omega_m$ is the magnon frequency, $\omega_p$ is the photon frequency, and $g$ is the magnon-photon coupling strength due to dipolar interaction. The field dependence of $\omega_p$ can be extracted from the linear extrapolation of the background, and the field dependence of $\omega_m$ can be obtained from the broad-band FMR spectrum. Fits to Eq. (\ref{eq00}) yield $g/2\pi=45$ MHz. Using the damping rates of $\kappa_p/2\pi=2.7$ MHz for the superconducting resonator at 5.5 K \textcolor{black}{and $\kappa_m/2\pi=50$ MHz for the Cu-EA acoustic mode, we obtain a cooperativity of $C=g^2/\kappa_p\kappa_m=15$. We note that even though the cooperativity becomes higher at lower temperature, e.g. 1.6 K, because of the much lower $\kappa_p$, the real bottleneck of strong magnon-photon coupling is the ratio $g/\kappa_m$, which is maximized as 0.95 at 5.5 K. The strong coupling regime requires both $g/\kappa_m$ and $g/\kappa_p$ to be greater than one \cite{ZhangPRL2014}. See the Supplemental Materials for the temperature dependence of $\kappa_p$ and $\kappa_m$ \cite{sm}.}

\begin{figure}[htb]
 \centering
 \includegraphics[width=3.0 in, angle = 0]{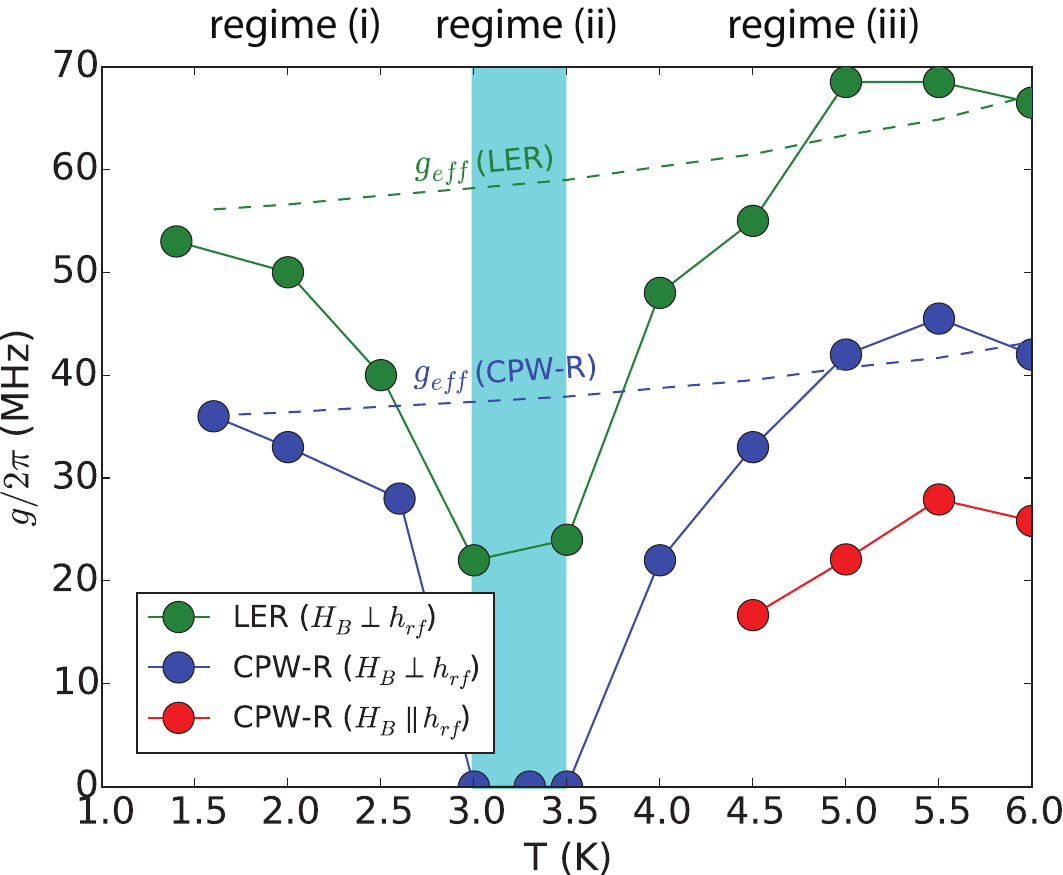}
 \caption{Extracted effective magnon-photon coupling $g$ as a function of $T$. The resonator mode is within the magnon band gap between 3 and 3.5 K, yielding $g=0$.}
 \label{fig3}
\end{figure}

Next, we investigate the temperature dependence of the magnon-photon interactions. Shown in Figs. \ref{fig2}(a) and (b), the signal line of the resonator generates both the in-plane and perpendicular Oersted fields, $h_\textrm{rf}^y$ and $h_\textrm{rf}^z$, respectively. \textcolor{black}{In the orthogonal pumping condition ($\mu_{0}H_B\perp h_\textrm{rf}^y$), the Oersted field components $h_\textrm{rf}^y$ and $h_\textrm{rf}^z$ only couple to the acoustic mode. In the parallel pumping condition ($\mu_{0}H_B\parallel h_\textrm{rf}^y$), $h_\textrm{rf}^y$ couples to the acoustic mode and $h_\textrm{rf}^z$ couples to the optical mode. Thus, the field alignment allows for selective excitation of the acoustic mode in Fig. \ref{fig2}(c), or the mutual excitation of both modes in Fig. \ref{fig2}(d).} The interaction with the acoustic mode is manifested by an avoided crossing at a constant field of $\mu_0H_\text{a}=95$ mT for both pumping geometries. The optical mode found in Fig. \ref{fig2}(d) shows a large temperature-dependent drift of its location, as marked by the dashed curves. The reversed anticrossing compared with the acoustic mode shows that the magnon frequency decreases as the field rises, agreeing with the feature of the optical mode as shown in Fig. \ref{fig1}(d).

Due to the tunability of $H_E$, the center frequency of the magnon band gap changes rapidly with temperature in the range from 1.5 to 6.0 K, therefore, allowing the resonator mode (much less sensitive to temperature) to intercept with the magnon band gap while maintaining a nearly constant quality factor. Three regimes of the magnon-photon coupling between the Cu-EA flake and the superconducting resonator are observed, with the relation between the magnon band gap and the resonator mode shown in Figs. \ref{fig2}(e-g). In regime (i) ($T>3.5$ K), the magnon band gap is below the superconducting resonator frequency [Fig. \ref{fig2}(e)]. The resonator photon mode \textcolor{black}{coherently} interacts with the acoustic mode in Fig. \ref{fig2}(c), and both the acoustic and optical modes in Fig. \ref{fig2}(d). In regime (ii) ($3.5\geq T \geq 3$ K), the acoustic and optical magnon modes cross each other and form the magnon band gap at the superconducting resonator mode frequency. This causes the resonator mode to fall inside the magnon band gap, leading to a moderate change of the peak amplitude and linewidth without peak frequency shift around 95 mT. \textcolor{black}{In regime (iii) ($T<3$ K) where the magnon band gap is above the resonator mode, the acoustic mode resumes its anticrossing-like interaction with the resonator mode. In addition, for the $\mu_{0}H_B\parallel h_\textrm{rf}^y$ geometry where the optical mode also interacts with the resonator mode [Fig. \ref{fig2}(d)], the regime between the optical and acoustic modes are blurred, as shown in regimes (i) and (iii). This indicates that one of the two acoustic-optical hybrid magnonic modes is still near the resonator mode and maintains the magnon-photon interaction.}

\begin{figure}[htb]
 \centering
 \includegraphics[width=3 in]{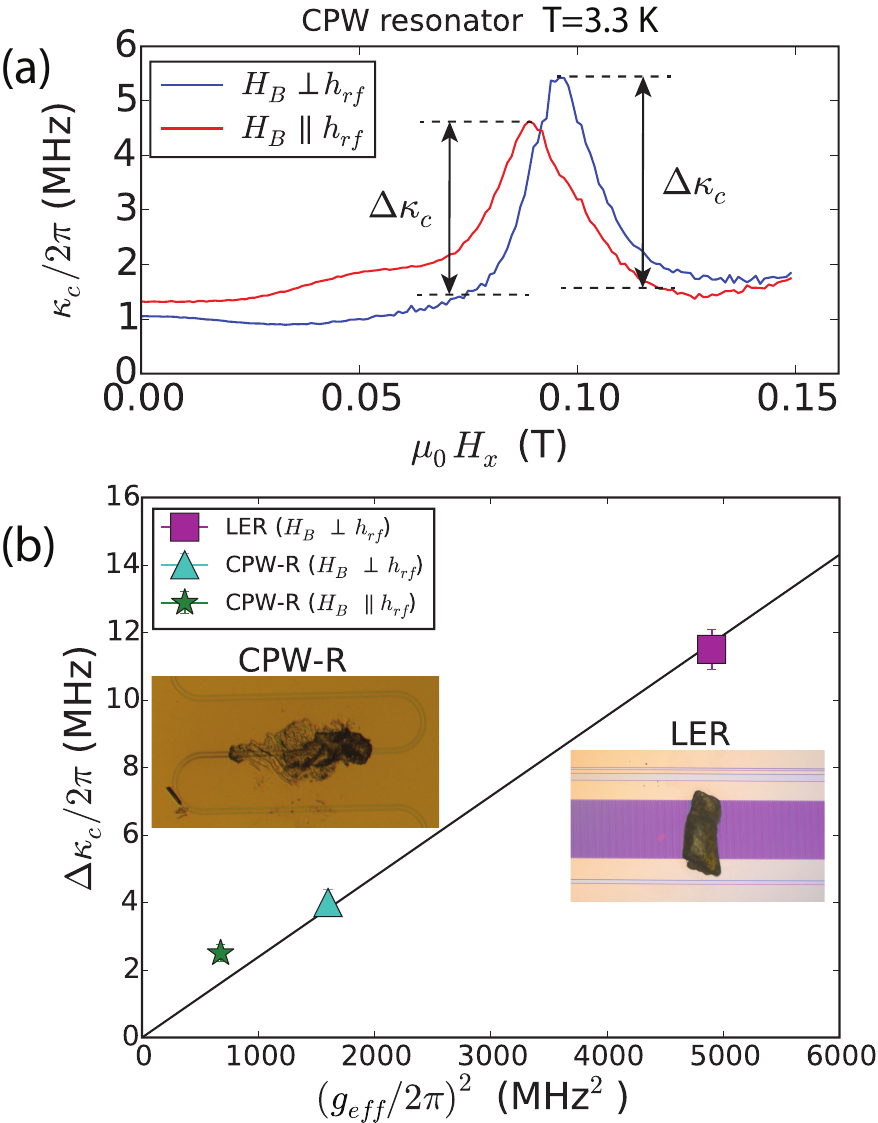}
 \caption{(a) Superconducting resonator linewidth $\kappa_c$ as a function of $H_B$ at $T=3.3$ K, where the resonator mode is inside the magnon band gap. (b) SC resonator linewidth change $\Delta\kappa_c$ as a function of $g_\text{eff}^2$ for the CPW and LER resonator designs. Error bars denote the uncertainty of base resonator linewidth drift under external magnetic fields. The red line is a fit to Eq. \ref{eq01}, with the slope quantifying the magnon band gap $2\delta$.}
 \label{fig4}
\end{figure}

Figure \ref{fig3} summarizes extracted magnon-photon coupling strength, $g$, as a function of $T$. To verify the phenomena, we have also coupled another Cu-EA crystal to a lumped-element resonator (LER) \cite{McKenziePRB2019}. This allows for a larger magnon-photon coupling strength while maintaining the same magnon band gap. For both the CPW resonator and LER, $g$ quickly decreases in regime (ii) due to mode degeneracy breaking between the magnon mode and the resonator photon mode. The zero coupling strength for the CPW resonator is manifested by the continuous evolution of resonator peak without mode anticrossing, as shown in Fig. \ref{fig2}(c-d). For the LER, a finite $g$ can still be extracted in regime (ii) which is due to non-perfect centering of the resonator mode in the magnon band gap when then magnon-photon coupling is large. The maximal acoustic mode coupling strengths for the CPW resonator are $g_{\text{CPW}}^\perp/2\pi = 45$ MHz for $\mu_{0}H_B\perp h_\textrm{rf}^y$ and $g_{\text{CPW}}^\parallel/2\pi = 28$ MHz for $\mu_{0}H_B\parallel h_\textrm{rf}^y$ at 5.5 K. Their difference quantifies the coupling ratio of the acoustic magnon mode between the in-plane ($h_{rf}^y$) and perpendicular($h_{rf}^z$) Oersted fields from the CPW: at $\mu_{0}H_B\perp h_\textrm{rf}$, both $h_{rf}^y$ and $h_{rf}^z$ couple to the acoustic mode, while at $\mu_{0}H_B\parallel h_\textrm{rf}$, only $h_{rf}^z$ couples to the acoustic mode. The ratio can be calculated as $h_{rf}^y/h_{rf}^z=\sqrt{(g_{\text{CPW}}^\perp)^2-(g_{\text{CPW}}^\parallel)^2}/g_{\text{CPW}}^\perp=1.25$. For the LER, the obtained ratio is 1.23. This suggests that $h_{rf}^z$ plays an important role in magnon-photon coupling. When the magnon band gap is far from the resonator mode (e.g. 1.5 K and 6 K), a reduction of $g$ from 6 K to 1.5 K reflects the change of coupling efficiency between the Oersted field and the canted magnetization at different biasing field directions. We plot the calculated prediction of the effective magnon-photon coupling, $g_\text{eff}$, for the acoustic magnon mode without considering the magnon band gap \cite{sm}, and the trend nicely captures the experiment at low and high temperatures.

We show that the magnon band gap of Cu-EA can be quantitatively extracted from the modulated magnon-photon interaction. When the resonator mode is inside the magnon band gap in regime (ii), the interaction between the magnon and photon modes leads to a linewidth broadening of the resonator photon mode. Such an effect has been previously observed in magnon-magnon coupled bilayers in the Purcell regime \cite{XiongSREP20,XiongPRApplied22,InmanPApplied22,KhanPRB21}. We develop an analytical model for quantifying the change of photon linewidth by considering two detuned magnon mode coupled to the photon mode. The photon damping rate $\kappa_c$ can be expressed as:
\begin{equation}\label{eq01}
  \kappa_c = \kappa_{c0} + (g_\text{eff})^2{\kappa_m \over \kappa_m^2+\delta^2},
\end{equation}
where $\kappa_{c0}$ is the intrinsic photon damping rate, $g_\text{eff}$ is the effective magnon-photon coupling strength as plotted in Fig. \ref{fig3}, $\kappa_m$ is the magnon damping rate, and $2\delta$ is the magnon-magnon band gap. The detailed derivation of the model is included in the Supplemental Materials \cite{sm}. Note that the information of $g_\text{eff}$ needs to be obtained from regime (iii) where mode anticrossing between the magnon and photon modes are resumed. Eq. (\ref{eq01}) shows that the change of linewidth $\Delta\kappa_c=\kappa_c - \kappa_{c0}$ is proportional to $(g_\text{eff})^2$, with the slope determined by two intrinsic magnon characteristics of the Cu-EA: $\kappa_m$ and $2\delta$. With two completely different superconducting resonator designs, i.e., the CPW resonator and LER, we find that the extracted $\Delta\kappa_c$ nicely follows the linear dependence of $(g_\text{eff})^2$, with a slope of (210 MHz)$^{-1}$. \textcolor{black}{For $\kappa$, we take the average of the acoustic and optical modes, as $\kappa_m/2\pi=65$ MHz. The magnon band gap is calculated to be $\delta/2\pi=152$ MHz, which is close to the value in Fig. \ref{fig1}(e) as 140 MHz around 3.5 K.} Thus, we confirm the validity of this new technique for quantifying the magnon band gap $\delta$ of a small magnetic flake with a highly sensitive superconducting microwave resonator, where the linewidth change of the resonator mode acts as a probe to interact with the acoustic-optical hybrid magnon modes.

In summary, we demonstrate tunable magnon-photon coupling by adjusting the intrinsic magnon band gap in a layered perovskite antiferromagnet in coupling with a superconducting resonator. The use of high-quality-factor superconducting resonator allows for coherent interaction with the magnon excitations and the study of unique magnon band gap in a magnetic material with narrow-band microwave measurements. The magnon-photon coupling strength can be tuned from a few tens of megahertz to zero by modifying the magnon band gap location with temperature. At the zero coupling strength state where the resonator mode falls into the magnon band gap, probing the change of photon mode linewidth also allows one to extract the value of magnon band gap using an analytical model. \textcolor{black}{Our results provide a new idea to modify magnon-photon interaction as well as a new approach to study the quantum properties of novel layered magnetic materials from cavity magnonics.} \textcolor{black}{To improve the slow temperature tunability of magnon band gap, we anticipate other approaches such as strain or electric field \cite{WangAdvMater2018,YanAdvMater2020,LiuSciAdv21} for controlling the magnetic properties with high speed and extending the application in coherent information processing.}

\textbf{Acknowledgement.} D.S. and M.B. acknowledge the primary financial support through the Center for Hybrid Organic Inorganic Semiconductors for Energy (CHOISE), an Energy Frontier Research Center funded by the Office of Basic Energy Sciences, Office of Science within the U.S. Department of Energy (Hybrid perovskite synthesis, crystal preparation, structural characterization, and motivation of this work). This work was authored in part by the National Renewable Energy Laboratory (NREL), operated by Alliance for Sustainable Energy LLC, for the U.S. Department of Energy (DOE) under contract no. DE-AC36-08GO28308. The views expressed in this article do not necessarily represent the views of the DOE or the U.S. Government. Works at Argonne National Laboratory and University of Illinois and Urbana-Champaign, including the superconducting resonator fabrication, design, ICP chemical analysis, and hybrid magnonics characterization, were supported by the U.S. DOE, Office of Science, Basic Energy Sciences, Materials Sciences and Engineering Division under contract No. DE-SC0022060. Work at UNC-CH were supported by NSF-ECCS 2246254 for the experimental design, data analysis, theoretical analysis, and manuscript preparation. D.S. acknowledges the partial financial support from the Department of Energy grant DE-SC0020992 and the National Science Foundation grant DMR-2143642 for magnetic properties characterization. X.Z. acknowledges support by the National Science Foundation CAREER Award (Grant CBET-2145417) and LEAPS Award (Grant DMR-2137883) for the Raman characterization. Use of the Center for Nanoscale Materials (CNM), an Office of Science user facility, was supported by the U.S. Department of Energy, Office of Science, Office of Basic Energy Sciences, under Contract No. DE-AC02-06CH11357.


%

\end{document}